# Size-selective optical forces for microspheres using evanescent wave excitation of whispering gallery modes

Jack Ng and C.T. Chan

*Department of Physics, Hong Kong University of Science and Technology*

*Clear Water Bay, Hong Kong.*

We show that when a microsphere is illuminated by an evanescent wave, the optical forces on- and off- whispering gallery mode (WGM) resonance can differ by several orders of magnitude. Such size selective force allows one to selectively manipulate the resonating particles, while leaving those particles at off-resonance untouched. As WGM resonances have very high-$Q$'s, this kind of force could be deployed for size-selective manipulation with a very high accuracy (~$1/Q$), as well as simultaneous particle-sorting according to their size or resonant frequency.

By using an intense laser beam, it is well-known that one can trap or manipulate microscopic particles. For the same laser field, particles with different morphologies would experience different optical forces.[1,2,3] In this paper, we consider a type of size-selective optical force that can be achieved by utilizing evanescent waves to excite a microsphere's high-$Q$ whispering gallery mode (WGM). We will call this kind of morphology dependent optical force $F_{WGM}$.

There were already experimental observations[4,5,6,7] of light induced mechanical effects caused by WGMs. While WGM induced optical forces have been considered in previous experimental[4,5] and theoretical works,[8] the use of propagating waves show only modest contrast (~50%) in the optical force at on- and off- resonance. As a result, the forces at on- and off- resonance are of the same order of magnitude, hence size-selective manipulation is not particularly effective. We showed that an evanescent wave can induce a very large contrast between the optical force at on- and off- resonance, and hence allowing size-selective manipulation. In a collection of microspheres, an evanescent wave will exert significant forces on those microspheres whose sizes are in resonance with the incident light, while leaving those microspheres that are not at resonance alone. By doing so, one can achieve size-selective manipulation with a size-selectivity of $\sim 1/Q$, and one may also sort the microspheres according to their size or resonant frequency. Ultra high-$Q$ microsphere can now be routinely fabricated:[9] a lower bound of $Q \approx 10^4$ was reported for 5 µm-diam polystyrene spheres stained with dye.[10] Such high-$Q$ values give extremely size-sensitive $F_{WGM}$.

While there are other approaches that can determine a microsphere's size to an accuracy of $\sim 1/Q$ using WGM,[11,12,13] these approaches are neither automatic nor parallel, and therefore they have low throughputs. With $F_{WGM}$, one will be able to pick up microspheres with the desired resonant frequency and potentially in a large quantity. A convenient method to select particle is highly desirable as it is very difficult to accurately control the resonant frequencies of a microsphere during its fabrication. The WGM microcavities are important components in nanophotonics, quantum optics, nonlinear optics, and many other areas.[9] The ability to massively produce microsphere-cavities with nearly identical resonant frequency may open up new possibilities and applications. In the current state-of-the-art, the difficulties in collecting identical microspheres limit the experiments on tight binding photonic modes to "photonic molecules" consisting of a small number of particles.[10,11,12,13,14,15] The sorting scheme we proposed may help to collect a larger number of microspheres with size-dispersion fulfilling the stringent optical requirements, paving the way to go beyond small "photonic molecules" to extended photonic crystals, coupled resonator optical waveguides, and other structures.

We consider a microsphere illuminated by an evanescent wave.[16,17,18,19,20] The geometry of the problem is depicted in Fig. 1(c). The origin of the coordinate system lies at the center of the sphere. The evanescent wave is assumed to be decaying exponentially from an interface (for example, due to total internal reflection), and above the interface it has the form

$$\vec{E}_{in} = \begin{pmatrix} e^{-k(r_s+\lambda/2)\sqrt{k_\parallel^2/k^2-1}} \sqrt{2I_0/\varepsilon_0 c} \\ \times \exp\left\{ik_\parallel x - \sqrt{k_\parallel^2 - k^2}\, y\right\} \end{pmatrix} \begin{bmatrix} -p\sqrt{1-k_\parallel^2/k^2} \\ pk_\parallel/k \\ -s \end{bmatrix}, \qquad (1)$$

where $r_s$ is the radius of the sphere, $\lambda$ is the incident wavelength, $k$ is the wavenumber, $k_\parallel$ is the component of the wavevector parallel to the wave's propagating direction, and $s$ and $p$ are the coefficients of the $s$- and $p$- polarizations respectively. The edge of the sphere is assumed to be $\lambda/2$ away from the interface. We will assume that light intensity at the interface is $I_0 = 10^4$ W/cm$^2$, which can be achieved by focusing a 1 Watt laser to an area of 0.01 mm$^2$. Such an area is large for microscopic particles; consequently parallel manipulation is possible. We note that the optical force linearly scales with the incident intensity. In real implementation, the incident field can be further enhanced by coating the interface with a dielectric cavity layer,[21] or by coating the interface with plasmonic material.[22,23] We neglect the multiple scattering between the interface and the microspheres, which will reduce the $Q$. The influence of the interface on the microspheres' $Q$ will be accounted for by adding a small imaginary part to the dielectric constant of the sphere to adjust its $Q$.

We apply the multiple scattering and Maxwell stress tensor (MS-MST) formalism that we developed earlier to calculate the optical forces.[24,25] The incident wave is expanded as

$$\vec{E}_{in} = -\sum_{n=1}^{\infty}\sum_{m=-n}^{n} iE_{mn} \times \left[ p_{mn}\vec{N}_{mn}^{(1)} + q_{mn}\vec{M}_{mn}^{(1)} \right], \quad (2)$$

where $E_{mn} = \sqrt{2I_0/\varepsilon_0 c}(2n+1)i^n(n-m)!/(n+m)!$, $\vec{N}_{mn}^{(1)}$ and $\vec{M}_{mn}^{(1)}$ are the vector spherical harmonics, and $p_{mn}$ and $q_{mn}$ are the expansion coefficients. In the case of $s$- and $p$-polarization, the expansion coefficients are

$$\begin{aligned} p_{mn}^{s-pol} &= iq_{mn}^{p-pol} = \\ &\frac{E_0}{n(n+1)}\frac{2^{n+1}(-1)^{n+1}}{(n-m)!}\frac{\Gamma\left(\frac{n-m+2}{2}\right)}{\Gamma\left(\frac{-n-m}{2}\right)}f_{m,kr_s}(k_\parallel), \\ q_{mn}^{s-pol} &= ip_{mn}^{p-pol} = \\ &\frac{E_0 m}{n(n+1)}\frac{2^n(-1)^n}{(n-m)!}\frac{\Gamma\left(\frac{n-m+1}{2}\right)}{\Gamma\left(\frac{-n-m+1}{2}\right)}f_{m,kr_s}(k_\parallel), \end{aligned} \quad (3)$$

where

$$f_{m,kr_s}(k_\parallel) = \left(k_\parallel/k + \sqrt{k_\parallel^2/k^2 - 1}\right)^m e^{-k(r_s+\lambda/2)\sqrt{k_\parallel^2/k^2-1}} \quad (4)$$

contains the $k_\parallel$-dependence of the expansion coefficients.

The natural modes of a sphere are characterized by the angular ($n$), azimuthal ($m$), radial ($r$), and polarization (TE or TM) numbers. We shall label the modes as ($n$)TE($r$) or ($n$)TM($r$), with TE and TM denote transverse electric and transverse magnetic respectively. Modes with different $m$'s are degenerate for a sphere. High $n$ and low $r$ modes are known as WGMs. We note that the range of (experimental) $Q$ for WGMs varies from ~$10^3$ to $10^{10}$. It is therefore possible for one to choose a resonance with an appropriate $Q$ that fits the application.

A typical radiation pressure spectrum as a function of the size parameter for a sphere in air under plane wave illumination is plotted in Fig. 1(a). The parameters are incident wavelength $\lambda = 520$ nm, dielectric constant $\varepsilon_{sphere} = 2.5281$ (~polystyrene), radius $r_s = 2.317 - 2.375$ $\mu$m (size parameter $kr_s = 28 - 28.7$), and incident intensity $I_0 = 10^4$ W/cm$^2$. The range of size parameter we consider is representative because it contains both TE and TM resonances, and from order 1 to 3. The optical force induced by an $s$-polarized evanescent wave is shown in Fig. 1(b). The parameters are the same as those in the plane wave case, except that the evanescent wave has an intensity $I_0$ at the interface where it is generated, and in order to mimic the sphere's intrinsic loss and the coupling loss with the interface, a small imaginary part is added to the dielectric constant ($\varepsilon_{sphere} = 2.5281 + 10^{-5} i$).[9,10,14] As expected from the Mie theory, all the resonant peaks in Fig. 1(a)-(b) have Lorentzian lineshapes, which implies that the magnitudes of the forces are directly proportional to the $Q$. The strongest peak in Fig. 1(b) is the 39TE1 mode. At the assumed loss level, $Q = 2.6 \times 10^5$ and $F_{WGM} = 6$ pN. The second strongest peak is the 35TE2 with $Q = 1.5 \times 10^4$ and $F_{WGM} = 0.5$ pN. For comparison the weight of a 5 μm-diam polystyrene sphere is ~ 0.7 pN; consequently pico-Newton-optical force can cause significant acceleration to the particle. The fact that the TE resonances are stronger than the TM resonances in Fig. 1(b) is a consequence of the polarization: if, *p*-polarization is used instead of *s*-polarization, the TM resonances will then be stronger. We note that if the actual loss of the microsphere-interface system is lower than the loss level we assumed, the force will be stronger than what is predicted in Fig. 1(b). In the case of a higher actual loss, one will have to use a stronger incident intensity to maintain the optical force to a level that is appropriate for optical manipulation.

Comparing Fig. 1(a) and Fig. 1(b), it can be seen that the contrast between the optical force at on- and off-resonance is significantly higher for the evanescent wave. For the case of an incident plane wave shown in Fig. 1(a), the optical force is always between ~2.5 to 5 pN, irrespective of whether a resonance is excited. On the contrary, for the case of an evanescent wave, the optical force at on- and off-resonance can differ by several orders of magnitude (note the log scale on Fig. 1(b)). The off-resonance optical force in the evanescent wave case is negligibly small (~fN), because the evanescent wave has a decaying amplitude, therefore the average amount of incident field that impinges the microsphere is small. However, when the WGM resonance is excited, the WGM couples with the evanescent wave very effectively such that the force is enhanced by several orders of magnitude in comparison with the off-resonance force, and leading to a sizable optical force. Furthermore, the linewidth of the WGM is narrow, which opens up the possibility of highly accurate size-selective manipulation.

We now explore the effect of finite laser linewidth $\Gamma_{Laser}$ on the attainable force for high-*Q* WGM. If the WGM's linewidth $\Gamma_{WGM}$ is narrower than $\Gamma_{Laser}$, only a fraction of the incident light is projected onto the WGM. Consequently, the strength of the force is decreased by a factor of $\sim \Gamma_{WGM}/\Gamma_{Laser}$. Nevertheless, sufficiently large optical force can be realized with commercially available lasers. Consider, for example, a laser operating at $\lambda = 520$ nm, $\Gamma_{Laser} \sim 0.025$ nm, and $I_0 = 10^4$ W/cm$^2$, the corresponding optical force for 39TE1 with $Q = 2.6\times10^5$ is ~0.5 pN.

Optical forces are typically categorized into two types: the nonconservative scattering force and the conservative gradient force. Fig. 1 plotted the scattering forces. For the evanescent wave excitation, there is also a strong resonant gradient force that acts on the sphere along $\hat{y}$, owing to the varying intensity of the evanescent wave. Both the scattering force $F_x$ and the gradient force $F_y$ for 39TE1 are plotted in Fig. 2(a). The scattering force is induced by strong scattering; hence it has a Lorentzian lineshape just as the scattering cross section. The gradient force can be attractive (red detuning) or repulsive (blue detuning) with respect to the source of the evanescent wave, and it is almost equally strong compared to the resonant scattering force. It might be counter-intuitive, at a first glance, to have a repulsive gradient force for a dielectric sphere, but the lineshape of the gradient force can be understood as follows. We note that near resonance, the resonating WGM dominates, and $F_{WGM}$ is a bilinear product of the incident driving field and the WGM's field.[26] When the size parameter changes across the resonance value, the field of the WGM changes from in-phase to out-of-phase with the driving field, and the sign of the force changes from negative to positive accordingly.

Fig. 2(b) shows the resonant scattering force $F_x$ versus $k_\parallel$, for both 39TE1 (square) and 35TE2 (triangle). As seen in Fig. 2(b), the optimal coupling of the 39TE1 and 35TE2 are, respectively, at $k_\parallel/k \approx 1.1$ and $k_\parallel/k \approx 1.25$. In Fig. 2(b), we plot the function $\left|f_{m=n,kr_s}(k_\parallel)\right|^2$ defined in (4) with the thin line and the thick line showing the cases of $n$=35 and $n$=39, respectively. The $\left|f_{m=n,kr_s}(k_\parallel)\right|^2$ shows that the overlap integral between a vector spherical harmonics corresponding to a WGM and an evanescent wave peaks at some particular $k_\parallel$, and if that particular $k_\parallel$ can be matched by the incident wave, strong resonance force can be excited. Indeed, it can be seen from Fig. 2(b) that the $k_\parallel$-dependence of $F_{WGM}$ is similar to that of $\left|f_{m=n,kr_s}(k_\parallel)\right|^2$.

In this letter, we have studied WGM-induced size-selective optical forces acting on microspheres. We showed that WGMs can be excited very effectively by evanescent waves, while preserving their high-$Q$. The force can be extremely size-selective (~1/$Q$), and as such, it allows for potentially accurate size-selective manipulation, as well as parallel particle-sorting according to their size or resonant frequency. We note in passing that $F_{WGM}$ can also be expected from other non-spherical high-$Q$ WGM cavities.

The support of HKUST3/06C is gratefully acknowledged. The Authors would like to thank A.W.O. Poon, Z.F. Lin, and P. Tong for their helpful discussions.

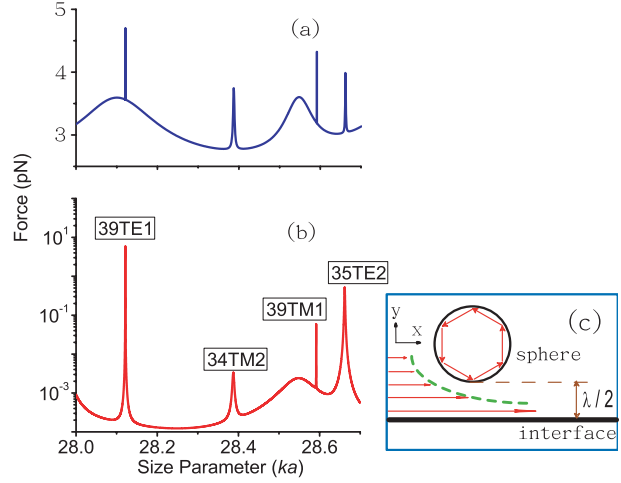

Fig. 1. (Color online) The optical force acting on a microsphere along the propagating direction of the incident wave. The incident wavelength has $\lambda = 520$ nm and the sphere radius $r_s = 2.317 - 2.375$ $\mu$m. **(a)** For the case $\varepsilon_{sphere} = 2.5281$. The incident wave is a linear polarized homogeneous plane wave with a uniform intensity of $I_0 = 10^4$ W/cm$^2$. **(b)** For $\varepsilon_{sphere} = 2.5281 + 10^5 i$. The incident wave is an *s*-polarized evanescent wave with $k_\parallel / k = 1.25$ and $I_0 = 10^4$ W/cm$^2$ at the interface where the evanescent wave is generated. **(c)** A schematic illustration of how the evanescent wave excites a WGM.

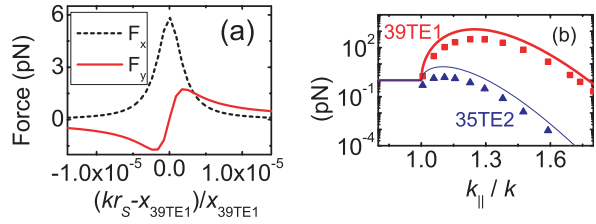

Fig. 2 (Color online) $F_{WGM}$ of 39TE1 and 35TE2 modes. The resonant size parameters are $x_{39TE1} = 28.1214244$ and $x_{35TE2} = 28.66183$. (a) The incident wave is an *s*-polarized evanescent wave with $k_\parallel / k = 1.25$ and $\varepsilon_{sphere} = 2.5281 + 10^5 i$. The scattering force $F_x$ of 39TE1 is shown as dotted line and the gradient force $F_y$ of 39TE1 is shown as solid line. (b) For the case of $\varepsilon_{sphere}=2.5281$. $F_x$ for 39TE1 and 35TE2 as a function of $k_\parallel$ are shown as square and triangle respectively. For comparison $|f_{n,ka}(k_\parallel)|^2$ is also plotted, with thick line for $n$=39 and thin line for $n$=35.